\documentclass[12pt, onecolumn, draftclsnofoot]{IEEEtran}

\usepackage[noadjust]{cite}
\usepackage{url}
\usepackage{amssymb, amsmath}

\newtheorem{theorem}{Theorem}
\newtheorem{lemma}[theorem]{Lemma}
\newtheorem{corollary}[theorem]{Corollary}
\newtheorem{definition}[theorem]{Definition}
\newtheorem{example}[theorem]{Example}

\newenvironment{remark}{\par {\itshape Remark:}}{}

\renewcommand{\QED}{\QEDopen}

\newcommand{\beqn}{\begin{equation}}
\newcommand{\eeqn}{\end{equation}}
\newcommand{\beq}{\begin{equation*}}
\newcommand{\eeq}{\end{equation*}}
\newcommand{\Z}{\mathbb Z}
\newcommand{\Cn}{\mathbb C}
\newcommand{\C}{{\cal C}}
\renewcommand{\L}{{\cal L}}

\renewcommand{\th}{{\rm th}}

\newcommand{\RM}{{\rm RM}}
\newcommand{\ZRM}{{\rm ZRM}}
\newcommand{\PMEPR}{\mbox{PMEPR}}
\newcommand{\PAPR}{\mbox{PAPR}}

\newcommand{\dist}{{\rm d}}

\newcommand{\bif}{{\boldsymbol{f}}}
\newcommand{\bix}{{\boldsymbol{x}}}
\newcommand{\bic}{{\boldsymbol{c}}}
\newcommand{\bid}{{\boldsymbol{d}}}
\newcommand{\biu}{{\boldsymbol{u}}}
\newcommand{\biv}{{\boldsymbol{v}}}
\newcommand{\biw}{{\boldsymbol{w}}}
\newcommand{\biy}{{\boldsymbol{y}}}

\newcommand{\biA}{{\boldsymbol{A}}}
\newcommand{\biB}{{\boldsymbol{B}}}
\newcommand{\biC}{{\boldsymbol{C}}}
\newcommand{\biD}{{\boldsymbol{D}}}
\newcommand{\biF}{{\boldsymbol{F}}}

\begin{document}

\title{On Cosets of the Generalized First-Order  Reed--Muller Code with Low PMEPR}%

\author{\large Kai-Uwe Schmidt\thanks{Kai-Uwe Schmidt is with Communications Laboratory, Dresden University of Technology, 01062 Dresden, Germany, e-mail: {schmidtk@ifn.et.tu-dresden.de}, web: {http://www.ifn.et.tu-dresden.de/$\sim$schmidtk/}}}%

\maketitle


\begin{abstract}
Golay sequences are well suited for use as codewords in orthogonal frequency-division multiplexing (OFDM) since their peak-to-mean envelope power ratio (PMEPR) in $q$-ary phase-shift keying (PSK) modulation is at most 2. It is known that a family of polyphase Golay sequences of length $2^m$ organizes in $m!/2$ cosets of a $q$-ary generalization of the first-order Reed--Muller code, $\RM_q(1,m)$. In this paper a more general construction technique for cosets of $\RM_q(1,m)$ with low PMEPR is established. These cosets contain so-called near-complementary sequences. The application of this theory is then illustrated by providing some construction examples. First, it is shown that the $m!/2$ cosets of $\RM_q(1,m)$ comprised of Golay sequences just arise as a special case. Second, further families of cosets of $\RM_q(1,m)$ with maximum PMEPR between 2 and 4 are presented, showing that some previously unexplained phenomena can now be understood within a unified framework. A lower bound on the PMEPR of cosets of $\RM_q(1,m)$ is proved as well, and it is demonstrated that the upper bound on the PMEPR is tight in many cases. Finally it is shown that all upper bounds on the PMEPR of cosets of $\RM_q(1,m)$ also hold for the peak-to-average power ratio (PAPR) under the Walsh--Hadamard transform.
\end{abstract}


\section{Introduction}

Despite many evident advantages of the orthogonal frequency-division multiplexing (OFDM) modulation technique, the widespread acceptance of OFDM mainly suffers from the usually high peak-to-mean envelope power ratio (PMEPR) of uncoded OFDM signals. A promising and elegant approach to solve this power control issue is to use a block code across the subcarriers \cite{Jones1994}. A well-designed code is able to provide a certain level of error protection and ensures a maximum PMEPR that is substantially reduced compared to uncoded transmission \cite{Jones1996}. 
\par
It has been proposed in a number of contributions \cite{Wilkinson1995}, \cite{Nee1996}, \cite{Ochiai1997} to use so-called \emph{Golay sequences} \cite{Golay1961} as codewords in OFDM since their PMEPR is at most 2 in $q$-ary phase-shift keying (PSK) modulation \cite{Popovic1991}. Major progress has been made by Davis and Jedwab \cite{Davis1999}; slightly generalized by Paterson \cite{Paterson2000a}, it was proved that a large family of polyphase Golay sequences of length $2^m$ organizes in $m!/2$ cosets of a naturally generalized first-order Reed--Muller code $\RM_q(1,m)$ inside a generalized second-order Reed--Muller code $\ZRM_q(2,m)$ (see Definition \ref{def:RM-codes}). A so-called supercode was then constructed by taking the union of these $m!/2$ cosets. However, due to a vanishing rate, these codes are only suitable for, say, $m\le5$. For this reason, Davis and Jedwab \cite{Davis1999} proposed to include further cosets of $\RM_q(1,m)$ with PMEPR greater than 2 (but still low), in order to increase the rate of the codes at the cost of a slightly larger PMEPR. An exhaustive search technique was employed to identify such cosets. However, due to high complexity, such a search becomes prohibitive when $m>4$ and $q$ increases.
\par
Paterson's work \cite{Paterson2000a} and the more general study by the author and Finger \cite{Schmidt2006b} provided some advanced theoretical background: it was shown that each coset of $\RM_q(1,m)$ is entirely comprised of sequences lying in so-called \emph{complementary sets} of the same size. It follows that the PMEPR of the codewords in each coset of $\RM_q(1,m)$ is at most $2^{k+1}$, where $k$ depends on the algebraic structure of a coset representative. However this upper bound is not always tight, and at best it yields merely the lowest possible power of 2.
\par
Further results were obtained by Parker and Tellambura. In \cite{Parker2003} they proposed another construction technique for cosets of $\RM_q(1,m)$ comprised of sequences lying in complementary sets of small size. In \cite{Parker2001b}, \cite{Parker2001} the Rudin--Shapiro construction \cite{Shapiro1951}, \cite{Rudin1959} was used to obtain sequence sets with low PMEPR from a suitable pair of starting sequences (kernel). However such a construction generally does not yield cosets of $\RM_q(1,m)$. In order to obtain complete cosets, an exhaustive search (although less complex than that proposed in \cite{Davis1999}) for suitable kernels was performed in \cite{Parker2001b} and \cite{Parker2001}. 
\par
Still, some phenomena cannot be fully explained. For example Davis and Jedwab observed \cite{Davis1999} that there are cosets of $\RM_8(1,m)$ whose maximum PMEPR, taken over all words in the coset, is equal to $3$. In addition numerical results suggest the existence of more coset classes whose PMEPR is bounded by low values not being a power of 2. This motivates further analyses of the PMEPR of cosets of $\RM_q(1,m)$. In particular it is of interest to identify cosets with PMEPR close to $2$.
\par
The remainder of this paper is structured as follows. In the next section we introduce the OFDM communication model and establish most of our notation. In Section \ref{sec:constructions} we use the Rudin--Shapiro construction to obtain sets of so-called \emph{near-complementary sequences} from a given kernel. This construction is then stated in Theorem \ref{thm:PMEPR_path} in the context of algebraic normal forms of generalized Boolean functions. These preliminary results serve as a stepping stone to establish our main result, summarized in Theorem \ref{thm:PMEPR_coset} and Corollary \ref{cor:PMEPR_cosets}, where we construct cosets of $\RM_q(1,m)$ from a kernel and prove an upper bound on the PMEPR of these cosets. In Section \ref{sec:lower-bounds} we prove a new lower bound on the PMEPR of arbitrary cosets of $\RM_q(1,m)$ and apply the results to those cosets constructed in Corollary \ref{cor:PMEPR_cosets}. These observations lead to some relations between the PMEPR and the peak-to-average power ratio (PAPR) under the Walsh--Hadamard transform. In Section \ref{sec:PAPR-WHT} we will comment on this issue and briefly discuss implications for the coding problem in multicode code-division multiple access (MC-CDMA) systems (cf. \cite{Paterson2002}, \cite{Paterson2004}).
\par
The application of our theory is then illustrated in Section \ref{sec:examples} by providing some construction examples. First, it is shown that the Davis--Jedwab construction of Golay sequences \cite{Davis1999} just arises as a special case in our theory, since they originate from trivial kernels (of length 1). Second, we present further classes of cosets of $\RM_q(1,m)$ whose maximum PMEPR is between 2 and 4. In particular we identify a class with PMEPR at most 3, and therefore, we provide a proof for the conjecture by Davis and Jedwab on the PMEPR of a subset of this class \cite{Davis1999}. The upper bound on the PMEPR is also compared with our lower bound, and it is shown that the upper bound is attained in many cases. Finally, in Section \ref{sec:conclusion}, we close with some concluding remarks and open problems.


\section{Notation and Preliminaries}

Throughout this paper $\xi=\exp(\sqrt{-1}\,2\pi/q)$ denotes a primitive $q\th$ root of unity and $q$ is an even positive integer. 

\subsection{Aperiodic Correlations and Complementarity}

Let $\biA,\biB\in\Cn^n$ be two sequences with $\biA=(A_0\,A_1\,\cdots\,A_{n-1})$ and $\biB=(B_0\,B_1\,\cdots\,B_{n-1})$. If $A_i=\xi^{a_i}$ and $a_i\in\Z_q$ for all $i=0,1,\cdots,n-1$, we shall call $\biA$ a \emph{polyphase sequence}. The \emph{aperiodic cross-correlation} of $\biA$ and $\biB$ at a displacement $\ell\in\Z$ is given by
\beq
C(\biA,\biB)(\ell)\triangleq 
\begin{cases}
\sum\limits_{i=0}^{n-\ell-1}A_{i+\ell}B^*_i&0\le \ell<n\\
\sum\limits_{i=0}^{n+\ell-1}A_iB^*_{i-\ell}&-n< \ell<0\\
0&\mbox{otherwise,}
\end{cases}
\eeq
where $()^*$ denotes complex conjugation. The \emph{aperiodic auto-correlation} of $\biA$ at a displacement $\ell\in\Z$ is then conveniently written as 
\beq
A(\biA)(\ell)\triangleq C(\biA,\biA)(\ell).
\eeq
\begin{definition}
\label{def:star}
For two sequences $\biA,\biB\in\Cn^n$ we define the operator '$\star$' as follows
\beq
\biA\star\biB\triangleq \sum_{\ell=1-n}^{n-1}|A(\biA)(\ell)+A(\biB)(\ell)|.
\eeq
\end{definition}
\vspace{1ex}
\par
If $\biA$ and $\biB$ are polyphase sequences of length $n$, we obtain
\beq
\biA\star\biB=2n+2\sum_{\ell=1}^{n-1}|A(\biA)(\ell)+A(\biB)(\ell)|.
\eeq
A pair of polyphase sequences $(\biA,\biB)$ is called a \emph{complementary pair} if $\biA\star\biB=2n$, which implies that the aperiodic auto-correlations of the two sequences sum up to zero for each nonzero shift. In tribute to Golay, who extensively studied binary complementary pairs in connection with multislit spectrometry \cite{Golay1961}, they are often called \emph{Golay complementary pairs}, and each sequence lying in such a pair is termed a \emph{Golay sequence}. 
\par
If $\biA$ and $\biB$ are polyphase sequences, $\biA\star\biB$ is at most $2n^2$. We shall call a pair of polyphase sequences $(\biA,\biB)$ a \emph{near-complementary pair} and the sequences therein \emph{near-complementary sequences} if $2n\le(\biA\star\biB)\ll 2n^2$. In other words, the aperiodic auto-correlations of $\biA$ and $\biB$ sum up to small values at a few nonzero shifts.

\subsection{OFDM Power Control}

Let us consider an $n$-subcarrier OFDM system. The transmitted OFDM signal is the real part of the \emph{complex envelope}, which can be written as
\beq
S(\biA)(\theta)\triangleq\sum\limits_{i=0}^{n-1}A_i\,e^{\,\sqrt{-1}\, 2\pi (i+\zeta) \theta},\quad 0\le \theta< 1,
\eeq
where $\zeta$ is a positive constant. The sequence $\biA=(A_0\,A_1\cdots A_{n-1})$ is called the modulating codeword of the OFDM symbol. Throughout this paper it is assumed that the elements of $\biA$ are selected from a PSK constellation. Then $\biA$ is a polyphase sequence.
\par
The PMEPR of the OFDM signal (or of the modulating codeword $\biA$) is then defined to be
\beqn
\label{eqn:PMEPR}
\PMEPR(\biA)\triangleq \frac{1}{n}\sup_{0\le \theta<1} |S(\biA)(\theta)|^2.
\eeqn
Notice that the PMEPR can be as large as $n$, which occurs, for example, if $\biA$ is the all-one word. However it is desirable to use codewords with PMEPR that is substantially lower than $n$. For the construction of such codewords the following theorem will be essential in the sequel.
\par
\begin{theorem}
\label{thm:PMEPR}
Let $(\biA,\biB)$ be a pair of polyphase sequences of length $n$. Then the PMEPR of $\biA$ and $\biB$ is at most $(\biA\star\biB)/n$.
\end{theorem}
\begin{proof}
It is well known (cf., e.g., \cite{Tellambura1997}, \cite{Davis1999}) that
\beq
|S(\biA)(\theta)|^2=A(\biA)(0)+2\sum\limits_{\ell=1}^{n-1}\Re\left\{A(\biA)(\ell)\,e^{\,\sqrt{-1}\, 2\pi \ell \theta}\right\},
\eeq
where $\Re\{.\}$ is the real part of a complex number. Hence
\begin{align*}
|S(\biA)(\theta)|^2+|S(\biB)(\theta)|^2&=2n+2\sum_{\ell=1}^{n-1}\Re\left\{\left[A(\biA)(\ell)+A(\biB)(\ell)\right]e^{\sqrt{-1} \,2\pi\ell \theta}\right\}\\
&\le 2n+2\sum_{\ell=1}^{n-1}\left|A(\biA)(\ell)+A(\biB)(\ell)\right|=\biA\star\biB,
\end{align*}
by Definition \ref{def:star}. Since $|S(.)(\theta)|^2$ of each individual sequence is non-negative, the PMEPR of $\biA$ and $\biB$ is at most $(\biA\star\biB)/n$.
\end{proof}
\par
The above theorem is consistent with the results in \cite{Popovic1991} in the special case where $(\biA,\biB)$ is a Golay complementary pair. Then the PMEPR of $\biA$ and $\biB$ is at most 2. In \cite{Wilkinson1995}, \cite{Nee1996}, \cite{Ochiai1997}, \cite{Davis1999} it has been proposed to exclusively use Golay sequences as codewords in OFDM. Consequently tight power control for OFDM is ensured, however, the code rate rapidly decreases for larger lengths. Theorem \ref{thm:PMEPR} motivates the use of larger sequence families with slightly higher PMEPR as codewords in OFDM.

\subsection{Generalized Boolean Functions and Associated Sequences}

A \emph{generalized Boolean function} $f$ is defined as a mapping $f\,:\,\Z_2^m\rightarrow\Z_q$. Such a function can be written uniquely in its \emph{algebraic normal form}, i.e., $f$ is the sum of $2^m$ weighted \emph{monomials}
\beq
f(\bix)=f(x_0,x_1,\cdots,x_{m-1})=\sum_{i=0}^{2^m-1}c_i\,\prod_{\alpha=0}^{m-1} x_\alpha^{i_\alpha},
\eeq
where the weights $c_0,\cdots,c_{2^m-1}$ are in $\Z_q$, and $(i_0\,i_1\cdots i_{m-1})$ is the binary representation of $0\le i<2^m$, such that $i=\sum_{j=0}^{m-1}i_j2^j$ is its binary expansion. The \emph{order of the $i\th$ monomial} is defined to be $\sum_{j=0}^{m-1}i_j$, and the \emph{order, or algebraic degree, of a generalized Boolean function} $f$, denoted by $\deg(f)$, is equal to the highest order of the monomials with a nonzero coefficient in the algebraic normal form of $f$. 
\par
A generalized Boolean function may be equally represented by sequences of length $2^m$. We shall define the sequence 
\beq
\psi(f)\triangleq (f_0\,f_1\cdots f_{2^m-1})
\eeq
as the \emph{$\Z_q$-valued sequence associated with $f$} and the sequence 
\beq
\Psi(f)\triangleq (\xi^{f_0}\,\xi^{f_1}\cdots \xi^{f_{2^m-1}})
\eeq 
as the \emph{polyphase sequence associated with $f$}. Here we denote $f_i=f(i_0,i_1,\cdots,i_{m-1})$, where $(i_0\,i_1\cdots i_{m-1})$ is the binary representation of $0\le i< 2^m$.
\par
In the remainder of this subsection we introduce the technique of extending polyphase sequences of length $2^m$ and their corresponding generalized Boolean functions. It will be used in the next section to prove our results on near-complementary sequences.
\par
\begin{definition}
\label{def:extended-vector}
Suppose $m>k$, let $f:\Z_2^k\rightarrow\Z_q$ be a generalized Boolean function in the variables $x_0,x_1,\cdots,x_{k-1}$, and write $\biF=\Psi(f)$. Let
\beq
0\le i_0<i_1<\cdots<i_{k-1}<m
\eeq
and write
\beq
0\le j_0<j_1<\cdots<j_{m-k-1}<m
\eeq
for the remaining indices. Also denote $\bix=(x_{j_0}\cdots x_{j_{m-k-1}})$ and let $\bid=(d_0\,d_1\cdots d_{m-k-1})$ be a binary word of length $m-k$. We define the \emph{extended sequence} $\biF_{[\bix=\bid]}$ of length $2^m$ as follows. As $(u_0\,u_1\cdots u_{k-1})$ ranges over $\Z_2^k$, at position
\beq
\sum_{\alpha=0}^{k-1}u_\alpha2^{i_\alpha}+\sum_{\alpha=0}^{m-k-1}d_\alpha2^{j_\alpha}
\eeq
the sequence $\biF_{[\bix=\bid]}$ is equal to $\xi^{f(u_0,u_1,\cdots,u_{k-1})}$ and equal to zero otherwise. We also define the \emph{extended generalized Boolean function} $f_{[\bix]}$ that is formally regarded as a generalized Boolean function in $m$ variables, i.e., it is of type $\Z_2^m\rightarrow\Z_q$. This function is obtained from $f$ by replacing the variables $x_\alpha$ by $x_{i_\alpha}$ in the algebraic normal form of $f$ for $\alpha=0,1,\cdots,k-1$.
\end{definition}
\par
Notice that $\biF_{[\bix=\bid]}$ comprises $2^k$ nonzero elements and $2^m-2^k$ zeros. It is a consequence of the above definition that at the positions where $\biF_{[\bix=\bid]}$ is nonzero the sequence $\biF_{[\bix=\bid]}$ is equal to the polyphase sequence associated with $f_{[\bix]}$.
\par
\begin{example}
Let $f:\Z_2^2\rightarrow\Z_2$ be given by
\beq
f=x_0x_1+x_1.
\eeq
Writing '$+$' for '$+1$' and '$-$' for '$-1$', we have $\biF=\Psi(f)=(+ + - +)$. Now take $\bix=(x_0\,x_2)$ and $\bid=(10)$. Then $\biF_{[\bix=\bid]}=(0\!+\!0\!+\!00000\!-\!0\!+\!0000)$. By relabeling the variable indices in $f$ according to $0\mapsto 1,1\mapsto 3$, we obtain the extended function $f_{[\bix]}=x_1x_3+x_3$. Regarding this function as a generalized Boolean function in $m$ variables, we obtain its associated polyphase sequence $(+ + + + + + + + - - + + - - + +)$. This sequence is equal to  $\biF_{[\bix=\bid]}$ at the positions where $\biF_{[\bix=\bid]}$ is nonzero.
\end{example}

\subsection{Generalized Reed--Muller Codes}

A code $\C$ of length $n$ over the ring $\Z_q$ is defined as a subset $\C\subseteq$ $\Z_q^n$. Such a code is called \emph{$\Z_q$-linear} if each $\Z_q$-linear combination of the codewords of $\C$ yields again a codeword of $\C$. If $\C$ is $\Z_q$-linear, a \emph{coset} of $\C$ is defined to be
\beq
\bif+\C\triangleq \{\bif+\bic\,|\,\bic\in\C\},
\eeq
where $\bif\in\Z_q^n$ is called its \emph{coset representative}, and of course, the additions are taken modulo $q$. We say that a coset of a code $\C_1$ lies inside a code $\C_2$ if $\bif+\C_1\subseteq\C_2$.
\par
We are interested in codes defined by generalized Boolean functions. In what follows we recall the definitions and some basic properties of the generalized Reed--Muller codes $\RM_q(r,m)$ and $\ZRM_q(r,m)$ (cf. \cite{Davis1999} and \cite{Paterson2000a}). 
\begin{definition} 
\label{def:RM-codes}
(a) For $0\le r\le m$ the code $\RM_q(r,m)$ is defined as the set of sequences $\psi(f)$, where $f$ is a generalized Boolean function $\Z_2^m\rightarrow\Z_q$ of order at most $r$.\\(b) For $q\ge 4$ and $1<r\le m$ the code $\ZRM_q(r,m)$ is defined as the set of sequences $\psi(f)$, where $f$ is a generalized Boolean function $\Z_2^m\rightarrow\Z_q$ with algebraic normal form containing monomials of order at most $r-1$ and two times the monomials of order $r$.
\end{definition}
\par
The codes $\RM_q(r,m)$ and $\ZRM_q(r,m)$ are $\Z_q$-linear, and their minimum Lee distances are equal to $2^{m-r}$ and $2^{m-r+1}$, respectively \cite{Davis1999}, \cite{Paterson2000a}. In this paper we will particularly study cosets of the code $\RM_q(1,m)$, which is comprised of the codewords corresponding to all affine forms over $\Z_q$ in $m$ two-state variables. Hence the number of words in such cosets is equal to $q^{m+1}$. 


\section{Constructions of Sequence Families with Low PMEPR}
\label{sec:constructions}

\subsection{Rudin--Shapiro Constructions}

In what follows we present a slightly generalized version of the Rudin--Shapiro construction \cite{Shapiro1951}, \cite{Rudin1959}, and exhibit its application to the construction of near-complementary pairs. Our main argument is the following lemma.
\begin{lemma}
\label{lem:rudin-shapiro}
Let $\biA$ and $\biB$ be two sequences of the same length and let $\biC=\biA+\biB$ and $\biD=\biA-\biB$. Then $\biC\star\biD=2(\biA\star\biB)$.
\end{lemma}
\par
\begin{proof}
It is straightforward to show
\begin{align*}
A(\biC)(\ell)&=A(\biA)(\ell)+A(\biB)(\ell)+C(\biA,\biB)(\ell)+C(\biB,\biA)(\ell)\\
A(\biD)(\ell)&=A(\biA)(\ell)+A(\biB)(\ell)-C(\biA,\biB)(\ell)-C(\biB,\biA)(\ell).
\end{align*}
Combining the relations above and Definition \ref{def:star}, the lemma follows.
\end{proof}
\par
It is well known that the Rudin--Shapiro construction can be employed to recursively construct longer complementary pairs starting from a known complementary pair \cite{Golay1961}, \cite{Budisin1990b}. Using the notation in the present paper, this will be illustrated in the context of a more general framework.
\par
Let $\biA$ and $\biB$ be two polyphase sequences of length $2^k$. The extended sequence $\biA_{[x_k=0]}$ is a sequence of length $2^{k+1}$ and contains the sequence $\biA$ in the left half, while its right half contains only zeros. Likewise the sequence $\biB_{[x_k=1]}$ contains zeros in the left half and the sequence $\biB$ in the right half. Observe that $A(\biA_{[x_k=0]})(\ell)=A(\biA)(\ell)$ and $A(\biB_{[x_k=1]})(\ell)=A(\biB)(\ell)$ for each $\ell\in\Z$. Let us construct
\begin{align*}
\biC&=\biA_{[x_k=0]}+\biB_{[x_k=1]}\\
\biD&=\biA_{[x_k=0]}-\biB_{[x_k=1]},
\end{align*}
and notice that the same pair could be constructed by
\beq
\biC=(\biA|\biB),\quad\biD=(\biA|{-\biB}),
\eeq
where $(.|.)$ denotes concatenation. The reader may recognize the classical Rudin--Shapiro construction. By Lemma \ref{lem:rudin-shapiro} we know that $\biC\star\biD=2(\biA\star\biB)$. Hence, if $(\biA,\biB)$ is a complementary pair, so will be $(\biC,\biD)$. This construction is also known as Golay's concatenation technique for synthesizing complementary pairs \cite{Golay1961}.
\par
We may also find a new sequence pair by
\begin{align*}
\biC&=\biA_{[x_0=0]}+\biB_{[x_0=1]}\\
\biD&=\biA_{[x_0=0]}-\biB_{[x_0=1]}.
\end{align*}
Notice that $\biA_{[x_0=0]}$ is obtained from $\biA$ by inserting zeros in $\biA$ at alternating positions starting at the second position. Likewise $\biB_{[x_0=1]}$ is obtained from $\biB$ by inserting zeros in $\biB$ at alternating positions starting at the first position. It is then easy to verify that $\biA_{[x_0=0]}\star\biB_{[x_0=1]}=\biA\star\biB$, and by Lemma \ref{lem:rudin-shapiro}, we have $\biC\star\biD=2(\biA\star\biB)$.  Therefore, if $(\biA,\biB)$ is a complementary pair, so will be $(\biC,\biD)$. This construction is essentially the same as Golay's interleaving technique for  synthesizing complementary pairs \cite{Golay1961}.
\par
If $\biA$ and $\biB$ are extended in more than one variable and Lemma \ref{lem:rudin-shapiro} is applied multiple times, then we can obtain even more general methods to construct longer sequence pairs from short ones. Again, if we restrict our attention to complementary pairs, we obtain Golay's general interleaving method to construct complementary pairs of length $2^m$ \cite{Golay1977}, \cite{Golay1961}, \cite{Budisin1990b}.
\par
We have now established the following. Starting from an initial polyphase sequence pair $(\biA,\biB)$ of length $2^k$ we can use the notion of extended sequences and a generalized Rudin--Shapiro construction to compose polyphase sequence pairs $(\biC,\biD)$ of length $2^m$ with $m>k$. Since by Theorem \ref{thm:PMEPR} we have $(\biC\star\biD)/2^m=(\biA\star\biB)/2^k$, the sequences $\biA$, $\biB$, $\biC$, and $\biD$ have the same PMEPR upper bound $(\biA\star\biB)/2^k$. If the sequence pair $(\biA,\biB)$ cannot be obtained from a shorter sequence pair in this way, then it is referred to as a \emph{kernel}.

\subsection{Explicit Constructions of Near-Complementary Sequences}

Using the language of generalized Boolean functions, an explicit construction for near-complementary sequences is presented in the following.
\begin{theorem}
\label{thm:PMEPR_path}
Let $m>k$ and write $m-k=s+t$ for non-negative integers $s,t$. Define the sets 
\beq
J=\{0,\cdots,s-1,m-t,\cdots,m-1\}
\eeq
and $I=\Z_m\backslash J$. Denote the elements of $J$ and $I$ by
\beq
0\le j_0<j_1<\cdots<j_{m-k-1}<m
\eeq
and
\beq
0\le i_0<i_1<\cdots<i_{k-1}<m,
\eeq
respectively.
Let $a,b:\Z_2^k\rightarrow\Z_q$ be two generalized Boolean functions and define $f:\Z_2^{m}\rightarrow\Z_q$ by
\begin{align*}
f(x_0,\cdots,x_{m-1})&=a(x_{i_0},\cdots,x_{i_{k-1}})(1-x_{j_{\pi(0)}})+b(x_{i_0},\cdots,x_{i_{k-1}})x_{j_{\pi(0)}}
\\
&\quad+\frac{q}{2}\sum_{\alpha=0}^{m-k-2}x_{j_{\pi(\alpha)}}x_{j_{\pi(\alpha+1)}}+\sum_{\alpha=0}^{m-k-1}w_\alpha x_{j_{\pi(\alpha)}}+w,
\end{align*}
where $w_0,\cdots,w_{m-k-1},w\in\Z_q$ and $\pi$ is a permutation of $\{0,1,\cdots,m-k-1\}$. Then 
\beq
\Psi(f)\star \Psi\left(f+\frac{q}{2}x_{j_{\pi(m-k-1)}}\right)=2^{m-k}\left[\Psi(a)\star\Psi(b)\right],
\eeq
and
\beq
\PMEPR(\Psi(f))\le\frac{\Psi(a)\star\Psi(b)}{2^k}.
\eeq
\end{theorem}
\vspace{1ex}
\par
\begin{proof}
Write $\bix=(x_{j_{\pi(0)}}\cdots x_{j_{\pi(m-k-1)}})$ and let $\bid=(d_0\cdots d_{m-k-1})$ be a binary word of length $m-k$. Moreover let $\biA=\Psi(a)$ and $\biB=\Psi(b)$. Consider the extended sequences $\biA_{[\bix=\bid]}$ and $\biB_{[\bix=\bid]}$. These sequences can be constructed by successively inserting zeros at alternating positions, at the beginning, or at the end of the sequences $\biA$ and $\biB$. It is then straightforward to establish that
\beqn
\biA_{[\bix=\bid]}\star\biB_{[\bix=\bid]}=\biA\star\biB.
\label{eqn:acf_extended}
\eeqn
\par
We now use a method similar to Golay's general interleaving construction, which is here applied to pairs that are not necessarily complementary. Let us define the recurrence formulae
\begin{align}
\label{eqn:ext1}
&\biC^{(\mu+1)}(d_{\mu+1},\cdots,d_{m-k-1})=
\biC^{(\mu)}(0,d_{\mu+1},\cdots,d_{m-k-1})+\xi^{w_\mu}\biD^{(\mu)}(1,d_{\mu+1},\cdots,d_{m-k-1})\\
\label{eqn:ext2}
&\biD^{(\mu+1)}(d_{\mu+1},\cdots,d_{m-k-1})=
\biC^{(\mu)}(0,d_{\mu+1},\cdots,d_{m-k-1})-\xi^{w_\mu}\biD^{(\mu)}(1,d_{\mu+1},\cdots,d_{m-k-1}),
\end{align}
where $\mu=0,1,\cdots,m-k-1$ and 
\begin{align*}
\biC^{(0)}(d_0,d_1,\cdots,d_{m-k-1})&=\biA_{[\bix=\bid]}\\ \biD^{(0)}(d_0,d_1,\cdots,d_{m-k-1})&=\biB_{[\bix=\bid]}.
\end{align*}
\sloppy
Since the positions of the nonzero components in $\biC^{(\mu)}(0,d_{\mu+1},\cdots,d_{m-k-1})$ and $\biD^{(\mu)}(1,d_{\mu+1},\cdots,d_{m-k-1})$ are distinct for each $\mu=0,\cdots,m-k-1$, $\biC^{(m-k)}$ and $\biD^{(m-k)}$ are polyphase sequences of length $2^m$. Observe that $A(\xi^{w_\mu}\biD^{(\mu)})(\ell)=A(\biD^{(\mu)})(\ell)$ for each $\ell\in\Z$. Therefore, by repeated application of Lemma \ref{lem:rudin-shapiro} and using (\ref{eqn:acf_extended}), we have 
\beqn
\label{eqn_CstarD}
\biC^{(m-k)}\star\biD^{(m-k)}=2^{m-k}(\biA\star\biB). 
\eeqn
\par
We know that $\biA_{[\bix=\bid]}$ and the polyphase sequence corresponding to $a_{[\bix]}$ are equal in those positions where $\biA_{[\bix=\bid]}$ is nonzero. An analogous statement holds for $\biB_{[\bix=\bid]}$ and $b_{[\bix]}$. Similarly we can find functions, say $c^{(\mu)}$ and $d^{(\mu)}$, whose associated polyphase sequences are equal to $\biC^{(\mu)}$ and $\biD^{(\mu)}$ at their respective nonzero components. Using (\ref{eqn:ext1}) and (\ref{eqn:ext2}) it can be verified that the functions $c^{(\mu)}$ and $d^{(\mu)}$ can be recursively constructed as follows
\begin{align*}
c^{(\mu+1)}&=c^{(\mu)}(1-x_{j_{\pi(\mu)}})+(d^{(\mu)}+w_{\mu})x_{j_{\pi(\mu)}}\\
d^{(\mu+1)}&=c^{(\mu+1)}+\frac{q}{2}x_{j_{\pi(\mu)}},
\end{align*}
where $c^{(0)}=a_{[\bix]}=a(x_{i_0},\cdots, x_{i_{k-1}})$ and $d^{(0)}=b_{[\bix]}=b(x_{i_0},\cdots, x_{i_{k-1}})$.
Explicitly we obtain
\begin{align*}
c^{(\mu)}&=a(x_{i_0},\cdots, x_{i_{k-1}})(1-x_{j_{\pi(0)}})+b(x_{i_0},\cdots, x_{i_{k-1}})x_{j_{\pi(0)}}\\
&\qquad\qquad+\frac{q}{2}\sum_{\alpha=0}^{\mu-2}x_{j_{\pi(\alpha)}}x_{j_{\pi(\alpha+1)}}+\sum_{\alpha=0}^{\mu-1}w_\alpha x_{j_{\pi(\alpha)}}\\
d^{(\mu)}&=c^{(\mu)}+\frac{q}{2}x_{j_{\pi(\mu-1)}},
\end{align*}
where $\mu=1,2,\cdots,m-k$. We have $f=c^{(m-k)}+w$ and $f+\frac{q}{2}x_{j_{\pi(m-k-1)}}=d^{(m-k)}+w$. The theorem follows then from (\ref{eqn_CstarD}) and Theorem \ref{thm:PMEPR}.
\end{proof}
\par
We refer to Section \ref{sec:conclusion} for a discussion on the relation of Theorem \ref{thm:PMEPR_path} to previous results in \cite{Parker2001b} and \cite{Parker2001}.

\subsection{Cosets of $\RM_q(1,m)$ with Low PMEPR}

Once suitable kernels are known, Theorem \ref{thm:PMEPR_path} identifies a large family of sequences with low PMEPR. However it would be desirable to construct sequence families that naturally form unions of cosets of $\RM_q(1,m)$ inside a higher-order generalized Reed--Muller code. In this way we could quickly obtain error-correcting codes, for which well-designed encoding and decoding algorithms exist (see, e.g., \cite{Davis1999}, \cite{Paterson2000b}, \cite{Grant1998}, \cite{Schmidt2006}, \cite{Schmidt2005a}). The remainder of this section is dedicated to finding such sequence sets. First we require some preliminaries.
\begin{definition}
\label{def:Phi}
Let $f:\Z_2^k\rightarrow\Z_q$ be a generalized Boolean function. We define the sequence $\Phi(f)$ of length $(4^k+2)/3$ as follows. As $(u_0\,u_1\cdots u_{k-1})$ ranges over $\Z_2^k$, at position 
\beq
\sum_{\alpha=0}^{k-1}u_\alpha2^{2\alpha}
\eeq
the sequence $\Phi(f)$ is equal to $\xi^{f(u_0,u_1,\cdots,u_{k-1})}$ and equal to zero otherwise.
\end{definition}
\par
We remark that the sequence $\Phi(f)$ may also be obtained from the extended sequence $\Psi(f)_{[\bix=\bid]}$ by setting $\bix=(x_1\,x_3\cdots x_{2k-3})$, letting $\bid$ be the all-zero word of length $k-1$, and deleting the trailing zeros. 
\par
Definition \ref{def:Phi} implies that for any generalized Boolean function $f:\Z_2^k\rightarrow\Z_q$ at its $2^k$ nonzero elements the sequence $\Phi(f)$ coincides with the polyphase sequence associated with a function that is obtained by replacing $x_\alpha$ by $x_{2\alpha}$ for each $\alpha=0,1,\cdots,k-1$ in the algebraic normal form of $f$, where this new function is regarded as the algebraic normal form of a generalized Boolean function in $2k-1$ variables.
\par
\begin{example}
Let $f:\Z_2^2\rightarrow\Z_4$ be given by
\beq
f(x_0,x_1)=2x_0x_1+3x_0+x_1.
\eeq
We obtain:
\begin{align*}
\Phi(f)&=(1\,-\!\!j\;0\;0\;j\,-\!\!1),\quad j=\sqrt{-1}.
\end{align*}
At the nonzero positions the above sequence coincides with the polyphase sequence associated with the function $g:\Z_2^3\rightarrow\Z_4$ whose algebraic normal form is given by $g(x_0,x_1,x_2)=2x_0x_2+3x_0+x_2$.
\end{example}
\par
We are now in the position to state the main theorem of this paper.
\begin{theorem}
\label{thm:PMEPR_coset}
Suppose that $m>k$. Let $a,b:\Z_2^k\rightarrow\Z_q$ be two generalized Boolean functions, and let $f:\Z_2^{m}\rightarrow\Z_q$, a generalized Boolean function in the variables $x_0,x_1,\cdots,x_{m-1}$, be given by
\begin{align*}
f(x_0,\cdots,x_{m-1})&=a(x_{\pi(0)},\cdots,x_{\pi(k-1)})(1-x_{\pi(k)})+b(x_{\pi(0)},\cdots,x_{\pi(k-1)})x_{\pi(k)}\\
&\quad+\frac{q}{2}\sum_{\alpha=k}^{m-2}x_{\pi(\alpha)}x_{\pi(\alpha+1)}+\sum_{\alpha=0}^{m-1}w_\alpha x_{\pi(\alpha)}+w,
\end{align*}
where $\pi$ is a permutation of $\{0,1,\cdots,m-1\}$ and $w_0,\cdots,w_{m-1},w\in\Z_q$. Then
\beq
\Psi(f)\star\Psi\left(f+\frac{q}{2}x_{\pi(m-1)}\right)\le2^{m-k}\left[\Phi(a)\star\Phi(b)\right]
\eeq
and
\beq
\PMEPR(\Psi(f))\le\frac{\Phi(a)\star\Phi(b)}{2^k}.
\eeq
\end{theorem}
\vspace{1ex}
\par
We need a series of lemmas in order to prove the theorem. It is well known that each integer $0\le i<2^m$ has a unique binary representation $(i_0\cdots i_{m-1})$, such that $i=\sum_{\alpha=0}^{m-1}i_\alpha2^\alpha$ and $i_\alpha\in\{0,1\}$. If we now allow $i_\alpha\in\{-1,0,+1\}$, we obtain a {\it signed-digit representation} (SDR) of $i$ \cite{Jedwab1989}. Such a representation is in general not unique. An SDR is called {\it sparse} if it contains no adjacent nonzero entries. We have the following lemma, which is a well-known result in number-representation theory (cf. \cite{Guentzer1987}, \cite{Jedwab1989}). A proof is included, since the lemma will play a crucial role in the sequel.
\begin{lemma}
\label{lem:sparse-SDR}
Every nonzero integer $i$ has a unique sparse SDR $(i_0\cdots i_{m-1})$ with $i_{m-1}\ne 0$.
\end{lemma}
\begin{proof} 
Without loss of generality we assume that $i>0$. We first show that for each nonzero integer there exists a sparse SDR. For $i=1$ this is obvious. The remaining cases are proved by induction. There are three cases: (i) if $i$ is even, take a sparse SDR for $i/2$ and prepend '$0$', (ii) if $i\equiv1\pmod 4$, take a sparse SDR for $(i-1)/4$ and prepend '$1,0$', (iii) if $i\equiv-1\pmod 4$, take a sparse SDR for $(i+1)/4$ and prepend '$-1,0$'.
\par
Now let us prove the uniqueness of a sparse SDR for a given positive integer. We first show that there is a unique sparse SDR for $i=1$ and then proceed by induction. Suppose $i_{m-1}=1$. Then
\beq
i\ge\left\{
\begin{array}{lll}
\displaystyle 2^{m-1}-\sum_{\alpha=0}^{(m-3)/2}2^{2\alpha}&\displaystyle \!\!\!\!=\frac{2^m+1}{3}&\mbox{if}\;\;m\;\;\mbox{is odd}\\
\displaystyle 2^{m-1}-\sum_{\alpha=0}^{(m-4)/2}2^{2\alpha+1}&\displaystyle \!\!\!\!=\frac{2^m+2}{3}&\mbox{if}\;\;m\;\;\mbox{is even.}
\end{array}
\right.
\eeq
If $i_{m-1}=-1$, a similar argument yields $i<0$. We conclude that, if $i=1$, then $m=1$ and $i_0=1$. Hence a sparse SDR for $i=1$ is unique.
\par
Now suppose that $i>1$ is the smallest positive integer with two sparse SDRs, namely $(u_0,\dots,u_{m-1})$ and $(v_0,\dots,v_{n-1})$. If $i$ is even, then $u_0=v_0=0$. By shifting the two SDRs one position to the left, we obtain two sparse SDRs for $i/2<i$, which is a contradiction. If $i\equiv 1\pmod 4$ or $i\equiv-1\pmod 4$, then $u_0=v_0=1$ or $u_0=v_0=-1$, respectively. Consequently, we would have two sparse SDRs for $(i-u_0)/4<i$. Again we arrive at a contradiction, which completes the proof.
\end{proof}
\par
We are now able to prove the following three lemmas.
\begin{lemma}
\label{lem:sum_acf}
Suppose $a,b:\Z_2^k\rightarrow\Z_q$ are two generalized Boolean functions. Then 
\beq
\Psi(a)_{[\bix=\bid]}\star\Psi(b)_{[\bix=\bid]}\le\Phi(a)\star\Phi(b)
\eeq
holds for any appropriate list of variables $\bix$ and for any $\bid$ of suitable length.
\end{lemma}
\par
\begin{proof}
Let 
\beq
0\le i_0<i_1<\cdots<i_{k-1}<m
\eeq
and write 
\beq
0\le j_0<j_1<\cdots<j_{m-k-1}<m
\eeq
for the remaining indices. Denote $\bix=(x_{j_0}\cdots x_{j_{m-k-1}})$ and let $\bid=(d_0\cdots d_{m-k-1})$ be a fixed binary word of length $m-k$. Write $\biA=\Psi(a)_{[\bix=\bid]}$, $\widetilde\biA=\Phi(a)$, $\biB=\Psi(b)_{[\bix=\bid]}$, and $\widetilde\biB=\Phi(b)$. We claim that
\beqn
\label{eqn:sum_acf}
A(\biA)(\ell)=\sum_{\ell'\in\L(\ell)}A(\widetilde\biA)(\ell')
\eeqn
for some disjoint sets $\L(\ell)$, which we shall now prove. Let $u$ and $u'$ have binary expansion
\begin{align*}
u&=\sum_{\alpha=0}^{k-1}u_\alpha2^{i_\alpha}+\sum_{\alpha=0}^{m-k-1}d_\alpha2^{j_\alpha}\\
u'&=\sum_{\alpha=0}^{k-1}u_\alpha2^{2\alpha},
\end{align*}
respectively, and use an analogous notation for $v$ and $v'$. Then $u$ and $u'$ define the positions of the nonzero components of $\biA$ and $\widetilde\biA$, respectively. By Definition \ref{def:extended-vector} and Definition \ref{def:Phi} we have $A_u=\tilde A_{u'}$ for any $u$ and corresponding $u'$. Now consider the nonzero product $\tilde A_{u'}\tilde A^*_{v'}=A_uA^*_v$ with $u'\ne v'$ (and therefore $u\ne v$) occuring in the expression $A(\widetilde\biA)(u'-v')$ and also in $A(\biA)(u-v)$. An SDR of $\ell'=u'-v'$ is given by
\beq
(u_0\!-v_0\;\;0\;\;u_1\!-v_1\,\cdots\,0\;\;u_{k-1}\!-v_{k-1}).
\eeq
This SDR is sparse and, by Lemma \ref{lem:sparse-SDR}, unique. In other words, for each $\alpha=0,\cdots,k-1$ the differences $u_\alpha-v_\alpha$ are uniquely determined by $\ell'$. Hence $\ell=u-v$ (which is independent of $\bid$) is also uniquely determined by $\ell'$. This means that all nonzero products contributing to the sum in the expression $A(\widetilde\biA)(\ell')$ also contribute to the sum in the expression $A(\biA)(\ell)$ for exactly one $\ell$, which proves (\ref{eqn:sum_acf}).
\par
Now we use (\ref{eqn:sum_acf}) to establish
\begin{align*}
\biA\star\biB&=2^{k+1}+2\sum_{\ell=1}^{2^m-1}|A(\biA)(\ell)+A(\biB)(\ell)|\\
&=2^{k+1}+2\sum_{\ell=1}^{2^m-1}\Bigg|\sum_{\ell'\in\L(\ell)}A(\widetilde\biA)(\ell')+\sum_{\ell'\in\L(\ell)}A(\widetilde\biB)(\ell')\Bigg|\\
&\le 2^{k+1}+2\sum_{\ell=1}^{2^m-1}\sum_{\ell'\in\L(\ell)}\big|A(\widetilde\biA)(\ell')+A(\widetilde\biB)(\ell')\big|.
\end{align*}
Since the sets $\L(\ell)$ are disjoint and each product contributing to $A(\widetilde\biA)(\ell')$ for some $\ell'$ also contributes to $A(\biA)(\ell)$ for some $\ell$, the latter expression is equal to $\widetilde\biA\star\widetilde\biB$, which completes the proof. 
\end{proof}
\par
\begin{lemma}
\label{lem:lin-terms}
Let $a,b:\Z_2^k\rightarrow\Z_q$ be two generalized Boolean functions, and define $\tilde a=a+L$ and $\tilde b=b+L$, where 
\beq
L=\sum_{\alpha=0}^{k-1}w_\alpha x_\alpha,\qquad w_0,\cdots,w_{k-1}\in\Z_q.
\eeq
Then
\beq
\Phi(\tilde a)\star\Phi(\tilde b)=\Phi(a)\star\Phi(b).
\eeq
\end{lemma}
\vspace{1ex}
\par
\begin{proof}
Write $\biA=\Phi(a)$, $\widetilde \biA=\Phi(\tilde a)$, $\biB=\Phi(b)$, and $\widetilde \biB=\Phi(\tilde b)$. Let $u$ and $v$ have binary representation $(u_0\cdots u_{2k-2})$ and $(v_0\cdots v_{2k-2})$, respectively. For $u\ne v$ we consider the products appearing in $A(\widetilde\biA)(u-v)$
\begin{align}
\tilde A_{u}\tilde A^*_{v}=&A_{u} \xi^{\sum_{\alpha=0}^{k-1}w_\alpha u_{2\alpha}}\,A^*_{v}\xi^{-\sum_{\alpha=0}^{k-1}w_\alpha v_{2\alpha}}\nonumber\\
=&A_{u}A^*_{v}\xi^{\sum_{\alpha=0}^{k-1}w_\alpha (u_{2\alpha}-v_{2\alpha})}.
\label{eqn:acf-lin}
\end{align}
Let us focus our attention on the word $(u_{0}-v_{0}\cdots u_{2k-2}-v_{2k-2})$, which is an SDR of $u-v$. By Definition \ref{def:Phi}, $A_u$ is equal to zero if and only if there exists an $\alpha\in\{0,\cdots,k-2\}$ such that $u_{2\alpha+1}\ne 0$. Therefore, if the product $A_uA^*_v$ is nonzero, an SDR of $u-v$ is sparse and, by Lemma \ref{lem:sparse-SDR}, unique. In this case the differences $u_{2\alpha}-v_{2\alpha}$ are uniquely determined by the difference $u-v$. Then we have $\tilde A_{u}\tilde A^*_{v}=A_{u} A^*_{v}\xi^{K(u-v)}$, where $K(u-v)$ only depends on $u-v$ and not explicitly on $u$ and $v$ themselves. Thus
\begin{align*}
A(\widetilde\biA)(\ell)&=\xi^{K(\ell)}A(\biA)(\ell),\\
\intertext{and similarly,}
A(\widetilde\biB)(\ell)&=\xi^{K(\ell)}A(\biB)(\ell).
\end{align*}
We conclude that
\beq |A(\widetilde\biA)(\ell)+A(\widetilde\biB)(\ell)|=|A(\biA)(\ell)+A(\biB)(\ell)|
\eeq
for all $\ell\in\Z$, and the lemma follows.
\end{proof}
\par
\begin{lemma}
\label{lem:permutations}
Let $a,b:\Z_2^k\rightarrow\Z_q$ be two generalized Boolean functions and define
\begin{align*} 
\tilde a(x_0,\cdots,x_{k-1})&=a(x_{\sigma(0)},\cdots,x_{\sigma(k-1)})\\
\tilde b(x_0,\cdots,x_{k-1})&=b(x_{\sigma(0)},\cdots,x_{\sigma(k-1)}),
\end{align*}
where $\sigma$ is a permutation of $\{0,1,\cdots,k-1\}$. Then
\beq
\Phi(\tilde a)\star\Phi(\tilde b)=\Phi(a)\star\Phi(b).
\eeq
\end{lemma}
\vspace{1ex}
\par
\begin{proof}
Write $\biA=\Phi(a)$, $\widetilde\biA=\Phi(\tilde a)$, $\biB=\Phi(b)$, and $\widetilde \biB=\Phi(\tilde b)$. Let $u$ and $u'$ have binary representation
\beq
(u_{0}\,u_{1}\,u_{2}\cdots u_{2k-3}\,u_{2k-2})
\eeq
and
\beq 
(u_{2\sigma(0)}\,u_{1}\,u_{2\sigma(1)}\cdots u_{2k-3}\,u_{2\sigma(k-1)}),
\eeq 
respectively. In an analogous manner we denote the binary representations of $v$ and $v'$. Let us consider the product $\tilde A_u\tilde A_v^*=A_{u'}A_{v'}^*$. By the same reasoning as in the proof of Lemma \ref{lem:lin-terms} we know that, if $A_{u'}A^*_{v'}$ is nonzero, an SDR of $u'-v'$ is sparse and uniquely determined by $u'-v'$. Similarly an SDR of $u-v$ is sparse if $\tilde A_u\tilde A_v^*$ is nonzero. It is then clear that, if $A_{u'}A_{v'}^*$ is nonzero, there is a one-to-one correspondence between the sparse SDR of $u-v$ and the sparse SDR of $u'-v'$, and hence, $u-v$ is uniquely determined by the difference $u'-v'$ and does not explicitly depend on $u'$ and $v'$ themselves. This means that each nonzero product appearing in $A(\biA)(u'-v')$ also appears in $A(\widetilde\biA)(u-v)$ and, by setting $\sigma:=\sigma^{-1}$ and arguing analogously, the converse holds too. Hence $A(\widetilde\biA)(u-v)=A(\biA)(u'-v')$ and, similarly, $A(\widetilde\biB)(u-v)=A(\biB)(u'-v')$. The lemma follows then immediately.
\end{proof}
\par
\noindent\hspace{2em}{\itshape Proof of Theorem~\ref{thm:PMEPR_coset}:}
Let a set of $m-k$ indices be given by $\{j_0,j_1,\cdots,j_{m-k-1}\}$, where $0\le j_\alpha<m$ for each $0\le\alpha<m-k$. Write $\bix=(x_{j_0}\cdots x_{j_{m-k-1}})$ and let $\bid$ be a binary word of length $m-k$. First we construct a number of initial function pairs $(\tilde a,\tilde b)$, where 
\begin{align*} 
\tilde a(x_0,\cdots,x_{k-1})&=a(x_{\sigma(0)},\cdots,x_{\sigma(k-1)})+L(x_0,\cdots,x_{k-1})\\
\tilde b(x_0,\cdots,x_{k-1})&=b(x_{\sigma(0)},\cdots,x_{\sigma(k-1)})+L(x_0,\cdots,x_{k-1}),
\end{align*}
$\sigma$ is a permutation of $\{0,1,\cdots,k-1\}$, and
\beq
L(x_0,\cdots,x_{k-1})=\sum_{\alpha=0}^{k-1}w_\alpha x_\alpha,\qquad w_0,\cdots,w_{k-1}\in\Z_q.
\eeq
From Lemma \ref{lem:sum_acf}, Lemma \ref{lem:lin-terms}, and Lemma \ref{lem:permutations} it follows
\beqn
\label{eqn:star_kernel}
\Psi(\tilde a)_{[\bix=\bid]}\star\Psi(\tilde b)_{[\bix=\bid]}\le\Phi(a)\star\Phi(b).
\eeqn
We can then proceed with the same constructive reasoning as in the proof of Theorem \ref{thm:PMEPR_path} to obtain the function pair $(c^{(m-k)},d^{(m-k)})$ from $(\tilde a,\tilde b)$. Using (\ref{eqn:star_kernel}) and the arguments from the proof of Theorem \ref{thm:PMEPR_path} we conclude
\beqn
\label{eqn:c_star_d}
\Psi(c^{(m-k)})\star \Psi(d^{(m-k)})\le 2^{m-k}[\Phi(a)\star \Phi(b)].
\eeqn
Then $c^{(m-k)}$ is a function where the addition of all linear terms in $m$ variables is possible. Moreover any permutation can be applied to the indices of the $m$ variables in $c^{(m-k)}$, because 
$\sigma$ and the indices $j_0,j_1,\cdots,j_{m-k-1}$ can be chosen arbitrarily, so that all $0\le j_\alpha<m$ are distinct. We can, therefore, replace $j_\alpha$ by $\pi(k+\alpha)$ for each $\alpha=0,1,\cdots,m-k-1$ and set $f=c^{(m-k)}+w$ and $f+\frac{q}{2}x_{\pi(m-1)}=d^{(m-k)}+w$. The theorem follows then from (\ref{eqn:c_star_d}) and Theorem \ref{thm:PMEPR}.
\hfill\QED
\par
We have a useful corollary of Theorem \ref{thm:PMEPR_coset}.
\begin{corollary}
\label{cor:PMEPR_cosets}
Let $m>k$ and suppose $a,b:\Z_2^k\rightarrow\Z_q$ are two generalized Boolean functions in $k$ variables. Then each polyphase codeword in the cosets $\psi(f)+\RM_q(1,m)$ with $f:\Z_2^m\rightarrow\Z_q$ given by
\begin{align*}
f=f(x_0,\cdots,x_{m-1})&=\frac{q}{2}\sum_{\alpha=k}^{m-2}x_{\pi(\alpha)}x_{\pi(\alpha+1)}\\
&\quad+a(x_{\pi(0)},\cdots ,x_{\pi(k-1)})(1-x_{\pi(k)})
+b(x_{\pi(0)},\cdots ,x_{\pi(k-1)})x_{\pi(k)},
\end{align*}
where $\pi$ is a permutation of $\{0,1,\cdots,m-1\}$, has PMEPR at most $[\Phi(a)\star\Phi(b)]/2^k$ and lies inside $\RM_q(r,m)$ with 
\beq
r=\begin{cases}
\max\{\deg(b-a)+1,\deg(a)\}& \mbox{if}\quad m=k+1\\
\max\{\deg(b-a)+1,\deg(a),2\}& \mbox{if}\quad m>k+1.
\end{cases}
\eeq
In particular, if $q\ge 4$ and all coefficients of the monomials in the algebraic normal form of $f$ with degree equal to $r$ are even, then the cosets are contained in $\ZRM_q(r,m)$. The number of distinct functions $f$, and therefore the number of distinct cosets, is at least $(m-k)!/2$ and at most $m!$ (the true value depends on how many permutations can be applied to the variable indices in the algebraic normal forms of $a$ and $b$ such that distinct pairs $(a,b)$ are generated).
\end{corollary}
\par
In order to construct families of cosets of $\RM_q(1,m)$ with low PMEPR, we just have to find two generalized Boolean functions $a,b:\Z_2^k\rightarrow\Z_q$ with $[\Phi(a)\star\Phi(b)]/2^k$ being low. This can be accomplished by an exhaustive search and, if $k$ is small enough, such functions could even be found using a 'by hand' construction technique. Several examples will be given in Section \ref{sec:examples}.


\section{Lower Bounds on the PMEPR}
\label{sec:lower-bounds}

Several lower bounds on the maximum PMEPR taken over all the words in a coset of $\RM_q(1,m)$ have been proposed in \cite{Paterson2000a}, \cite{Stinchcombe2000}, \cite{Manji2004}. These approaches rely on the examination of the OFDM signal at time $\theta=0$ or at some other $\theta$. Initially this method was proposed in \cite{Cammarano1999}. However existing results apply to second-order cosets of $\RM_q(1,m)$, where the coset representative is binary (in the $q$-ary context this means that it has values '$0$' and '$q/2$'). 
\par
In the following we provide a general lower bound on the PMEPR of the cosets $\psi(f)+\RM_q(1,m)$, where $f:\Z_2^m\rightarrow\Z_q$ is an arbitrary generalized Boolean function. A $q$-ary generalization of the Walsh--Hadamard transform plays an important role in our study. 
\begin{definition}
\label{def:WHT}
We define the $q$-ary Walsh--Hadamard transform (WHT) of a generalized Boolean function $f:\Z_2^m\rightarrow\Z_q$ (or of its polyphase sequence $\Psi(f)$) to be
\beq
F(\biw)\triangleq \sum_{\bix\in\Z_2^m}\xi^{f(\bix)+\biw \cdot \bix}, 
\eeq
where $\biw\in\Z_q^m$ and '$\cdot$' denotes the scalar product of vectors. By convention, in the special case where $m=0$ and $f=a$ ($a\in\Z_q$), we define $F$ to be equal to $\xi^a$.
\end{definition}
\par
If $q=2$, the above definition coincides with that of the classical WHT (see, e.g., \cite[Chapter 14]{MacWilliams1977}). The following lemma will be useful in the sequel.
\begin{lemma}
\label{lem:WHT-prop}
Let $f,g:\Z_2^m\rightarrow\Z_q$ be two generalized Boolean functions that are related by
\beq
g(\bix)=f(A\bix)+\biv\cdot\bix+v
\eeq
with $\biv\in\Z_q^m$, $v\in\Z_q$, and $A$ being an $m\times m$ permutation matrix. Then the sets $\{G(\biw)\,|\,\biw\in\Z_q^m\}$ and  $\{\xi^{v}F(\biw)\,|\,\biw\in\Z_q^m\}$ are equal.
\end{lemma}
\par
\begin{proof}
Write
\begin{align*}
G(\biw)&=\sum_{\bix\in\Z_2^m}\xi^{g(\bix)+\biw \cdot \bix}\\
&=\sum_{\bix\in\Z_2^m}\xi^{f(A\bix)+\biw \cdot \bix+\biv\cdot \bix+v}.
\end{align*}
Setting $\bix=A^{-1}\biy$, we obtain
\begin{align*}
G(\biw)&=\xi^v\sum_{\biy\in\Z_2^m}\xi^{f(\biy)+(\biw+\biv)\cdot (A^{-1}\biy)}\\
&=\xi^v\sum_{\biy\in\Z_2^m}\xi^{f(\biy)+(A(\biw+\biv))\cdot \biy}\\
&=\xi^{v}\,F(A(\biw+\biv)),
\end{align*}
and the lemma follows.
\end{proof}
\begin{remark}
In the particular case where $q=2$, Lemma \ref{lem:WHT-prop} still holds for $A$ being an arbitrary invertible binary matrix (cf. \cite[Chapter 14]{MacWilliams1977}).
\end{remark}
\par
Now we are ready to formulate the following theorem.
\begin{theorem}
\label{thm:lower-bound}
Let $f:\Z_2^m\rightarrow\Z_q$ be a generalized Boolean function. Then there exists a polyphase codeword in the coset  $\psi(f)+\RM_q(1,m)$ having PMEPR at least 
\beqn
\label{eqn:PAPR-WHT}
\frac{1}{2^m}\,\max_{\biw\in\Z_q^m}|F(\biw)|^2.
\eeqn
\end{theorem}
\vspace{1ex}
\par
\begin{proof}
Consider the coset $\psi(f)+\RM_q(1,m)$, which contains the polyphase codewords $\biF_{\biw w}=\Psi(f(\bix)+\biw\cdot\bix+w)$, where $\biw\in\Z_q^m$ and $w\in\Z_q$. The complex envelope corresponding to $\biF_{\biw w}$ reads for $\theta=0$
\begin{align*}
S({\biF_{\biw w}})(0)&=\sum_{i=0}^{2^m-1}F_{\biw w,i}\nonumber\\
&=\sum_{\bix\in\Z_2^m}\xi^{f(\bix)+\biw\cdot\bix+w}\\
&=F(\biw)\;\xi^w.
\end{align*}
Therefore
\begin{align*}
\max_{\biw,w}\sup_{0\le\theta<1}|S(\biF_{\biw w})(\theta)|^2
&\ge\max_{\biw,w}|S(\biF_{\biw w})(0)|^2\\
&=\max_{\biw,w}|F(\biw)\;\xi^w|^2\\
&=\max_{\biw}|F(\biw)|^2,
\end{align*}
and the theorem follows from (\ref{eqn:PMEPR}).
\end{proof}
\par
\begin{remark}
In the case where $f$ is a quadratic Boolean function and $q=2$, it can be shown that \cite[Theorem 1]{Manji2004} coincides with the above theorem.
\end{remark}
\par 
General lower bounds on the expression in (\ref{eqn:PAPR-WHT}) directly give lower bounds on the achievable PMEPR of cosets of $\RM_q(1,m)$. Let us first discuss the binary case, i.e., $q=2$. It is well known (cf., e.g., \cite[Chapter 14]{MacWilliams1977}) that
\beq
\max_{\biw\in\Z_2^m}|F(\biw)|=2^m-2\min_{\bic\in\RM_2(1,m)}\dist_H\left(\psi(f),\bic\right),
\eeq
where $\dist_H(\cdot,\cdot)$ is the Hamming distance between two binary sequences. The expression 
\beq
\rho(m)\triangleq \max_{\bif\in\Z_2^{2^m}}\min_{\bic\in\RM_2(1,m)}\dist_H\left(\bif,\bic\right)
\eeq
is the \emph{covering radius} of $\RM_2(1,m)$. Therefore, provided that $\rho(m)\le2^{m-1}$, we have
\beq
\min_f\max_{\biw\in\Z_2^m}|F(\biw)|=2^m-2\rho(m),
\eeq
where the minimum is taken over all Boolean functions $f:\Z_2^m\rightarrow\Z_2$. Results on the covering radius of $\RM_2(1,m)$  can now be used to lower-bound (\ref{eqn:PAPR-WHT}) for $q=2$ (cf., e.g., \cite{Brualdi1998}, \cite{Paterson2004}), and therefore, to lower bound the PMEPR of cosets of $\RM_2(1,m)$. If $m$ is even, we have $\rho(m)=2^{m-1}-2^{m/2-1}$, which leads to the trivial lower bound of $1$ for (\ref{eqn:PAPR-WHT}). For odd $m$ it is known that $\rho(m)\ge2^{m-1}-2^{(m-1)/2}$, where equality holds when $m\le7$. We conclude that the PMEPR of cosets of $\RM_2(1,m)$ is at least 2 if $m$ is odd and $m\le7$. For odd $m\ge 9$, the exact value of $\rho(m)$ is unknown. However, if $m\ge 15$, it is known that $\rho(m)\ge2^{m-1}-\frac{27}{32}2^{(m-1)/2}$, which implies that (\ref{eqn:PAPR-WHT}) can be strictly smaller than $2$ in this case.
\par
Much less can be said when $q>2$. However at least we can assert that the PMEPR of cosets of $\RM_q(1,m)$ must be strictly greater than 1 if $q$ is a multiple of 4. To this end observe
\beqn
\max_{\biw\in\Z_q^m}|F(\biw)|\ge\max_{\biw\in(q/4)\Z_4^m}|F(\biw)|.
\label{eqn:lower-bound-q4}
\eeqn
It can be shown that the set $\{F(\biw)\,|\,\biw\in\frac{q}{4}\Z_4\}$ is equal to the set of values of the $2^m$ possible $\{H,N\}^m$ transforms \cite{Riera2006} of a function $f:\Z_2^m\rightarrow\Z_q$. In \cite{Riera2006} it was proved that there is no function $f:\Z_2^m\rightarrow\Z_q$ for which all $2^m$ $\{H,N\}^m$ transforms contain only absolute values $2^{m/2}$. Moreover we conjecture that at least for small values of $m$ and for $q$ being a multiple of $4$ the right-hand side in (\ref{eqn:lower-bound-q4}) is lower-bounded by $2^{(m+1)/2}$, and therefore the PMEPR of cosets of $\RM_q(1,m)$ is at least $2$.
\par
In what follows we apply Theorem \ref{thm:lower-bound} to obtain lower bounds on the PMEPR of the cosets of $\RM_q(1,m)$ constructed in the previous section.
\begin{theorem}
\label{thm:LB-constr}
Let $a,b:\Z_2^k\rightarrow\Z_q$ be two generalized Boolean functions, and let $f:\Z_2^m\rightarrow\Z_q$ be given by
\begin{align*}
f(x_0,\cdots,x_{m-1})&=\frac{q}{2}\sum_{\alpha=k}^{m-2}x_{\pi(\alpha)}x_{\pi(\alpha+1)}\\
&\quad+a(x_{\pi(0)},\cdots ,x_{\pi(k-1)})(1-x_{\pi(k)})+b(x_{\pi(0)},\cdots ,x_{\pi(k-1)})x_{\pi(k)},
\end{align*}
where $\pi$ is a permutation of $\{0,1,\cdots,m-1\}$. Then there exists a polyphase codeword in the coset $\psi(f)+\RM_q(1,m)$
having PMEPR at least
\begin{align}
\label{eqn:lower_bound_m_even}
&\frac{1}{2^{k+2}}\max_{\biw\in\Z_q^{k+2}}\big|A(\biw')(1+\xi^{w_{k+1}})+B(\biw')\xi^{w_k}(1-\xi^{w_{k+1}})\big|^2\\
\intertext{if $m-k$ is even and at least}
\label{eqn:lower_bound_m_odd}
&\frac{1}{2^{k+1}}\max_{\biw\in\Z_q^{k+1}}\big|A(\biw')+B(\biw')\,\xi^{w_k}\big|^2
\end{align}
if $m-k$ is odd. Here $A(\biw')$ and $B(\biw')$ are the $q$-ary WHTs of $a$ and $b$, respectively, and $\biw'=(w_0\cdots w_{k-1})$.
\end{theorem}
\par
\begin{proof}
We intend to find a lower bound for
\beq
\frac{1}{2^m}\max_{\biw\in\Z_q^m}|F(\biw)|^2,
\eeq
where $F(\biw)$ is the $q$-ary WHT of $f$. Using Lemma \ref{lem:WHT-prop} we conclude that, in order to prove the theorem, it is sufficient to assume that $\pi$ is the identity permutation. So we are interested in the coset representative corresponding to 
\beq
f_m(\bix)=f_m(x_0,\cdots,x_{m-1})=\frac{q}{2}\sum_{\alpha=k}^{m-2}x_\alpha x_{\alpha+1}+a(x_0,\cdots,x_{k-1})(1-x_k)+b(x_0,\cdots,x_{k-1})x_k.
\eeq
Denote the $q$-ary WHT of $f_m$ by $F_m(\biw)$. We require the following expansion
\begin{align}
F_m(\biw)=&\sum_{\bix\in\Z_2^m}\xi^{f_m(\bix)+\biw\cdot\bix}\nonumber\\
=&\sum_{\biu\in\Z_2^{m-1}}\xi^{f_m(\biu,0)+\biv\cdot\biu}
+\xi^{w_{m-1}}\sum_{\biu\in\Z_2^{m-1}}\xi^{f_m(\biu,1)+\biv\cdot\biu},
\label{eqn:expand-WHT}
\end{align}
where $\biv=(w_0\cdots w_{m-2})$. Let us first consider the case $m=k+1$, so $m-k$ is odd. We have
\beq
f_{k+1}=a(x_0,\cdots,x_{k-1})(1-x_k)+b(x_0,\cdots,x_{k-1})x_k,
\eeq
and thus,
\begin{align*}
f_{k+1}(x_0,\cdots,x_{k-1},0)&=a(x_0,\cdots,x_{k-1})\\
f_{k+1}(x_0,\cdots,x_{k-1},1)&=b(x_0,\cdots,x_{k-1}).
\end{align*}
With (\ref{eqn:expand-WHT}) we obtain
\beqn
\label{eqn:WHT-k+1}
F_{k+1}(\biw',w_k)=A(\biw')+B(\biw')\xi^{w_k}.
\eeqn
Now let $m>k+1$. Then we have
\begin{align*}
f_m(x_0,\cdots,x_{m-2},0)&=f_{m-1}(x_0,\cdots,x_{m-2})\\
f_m(x_0,\cdots,x_{m-2},1)&=f_{m-1}(x_0,\cdots,x_{m-2})+\frac{q}{2}x_{m-2},
\end{align*}
and with (\ref{eqn:expand-WHT}) it follows that
\beqn
\label{eqn:WHT-induction}
F_m(w_0,\cdots,w_{m-1})=\,F_{m-1}(w_0,\cdots,w_{m-2})
+\,F_{m-1}(w_0,\cdots,w_{m-2}+\frac{q}{2})\xi^{w_{m-1}}.
\eeqn
For $m-k$ odd suppose 
\beqn
\label{eqn:F-odd}
F_m(\biw',w_k,0,\cdots,0)=2^{(m-k-1)/2}(A(\biw')+B(\biw')\xi^{{w_k}}),
\eeqn
which is true for $m=k+1$ (see (\ref{eqn:WHT-k+1})). We will use this expression as a hypothesis for the following induction. We employ (\ref{eqn:WHT-induction}) to obtain
\beqn
F_{m+1}(\biw',{w_k},0,\cdots,0,w_{m})
=2^{(m-k-1)/2}\left[A(\biw')(1+\xi^{{w_{m}}})+B(\biw')\xi^{{w_k}}(1-\xi^{{w_{m}}})\right]
\label{eqn:F-even}
\eeqn
and
\begin{align*}
F_{m+2}(\biw',w_k,0,\cdots,0,w_{m},w_{m+1})
&=2^{(m-k-1)/2}\left[A(\biw')(1+\xi^{{w_{m}}})+B(\biw')\xi^{{w_k}}(1-\xi^{{w_{m}}})\right.\\
&\quad\left.+\xi^{w_{m+1}}\left(A(\biw')(1-\xi^{{w_{m}}})+B(\biw')\xi^{{w_k}}(1+\xi^{{w_{m}}})\right)\right].
\end{align*}
Consequently we have
\begin{align*}
&F_{m+2}(\biw',w_k,0,\cdots,0)=2^{(m-k+1)/2}\left(A(\biw')+B(\biw')\xi^{{w_k}}\right),
\end{align*}
which proves by induction that (\ref{eqn:F-odd}) and (\ref{eqn:F-even}) hold in general. Now we can write for $m-k$ even
\begin{align*}
\frac{1}{2^m}\max_{\biw\in\Z_q^m}\big|F_m(\biw)\big|^2
&\ge\frac{1}{2^m}\max_{\biw\in\Z_q^{k+2}}\big|F_m(\biw',w_k,0,\cdots,0,w_{k+1})\big|^2\\
&=\frac{1}{2^{k+2}}\!\max_{\biw\in\Z_q^{k+2}}\!\big|A(\biw')(1+\xi^{{w_{k+1}}})+B(\biw')\xi^{{w_k}}(1-\xi^{{w_{k+1}}})\big|^2
\end{align*}
and for $m-k$ odd
\begin{align*}
\frac{1}{2^m}\max_{\biw\in\Z_q^m}\big|F_m(\biw)\big|^2
&\ge\frac{1}{2^m}\max_{\biw\in\Z_q^{k+1}}\big|F_m(\biw',w_k,0,\cdots,0)\big|^2\\
&=\frac{1}{2^{k+1}}\max_{\biw\in\Z_q^{k+1}}\big|A(\biw')+B(\biw')\xi^{{w_k}}\big|^2.
\end{align*}
Then the statements in the theorem follow from Theorem \ref{thm:lower-bound}.
\end{proof}


\section{The PAPR under the Walsh--Hadamard Transform}
\label{sec:PAPR-WHT}

Suppose that $f:\Z_2^m\rightarrow\Z_q$ is a generalized Boolean function and $p$ is a divisor of $q$. Then we define the \emph{peak-to-average power ratio} (PAPR) of $\Psi(f)$ under the $p$-ary WHT to be
\beq
\PAPR_p(\Psi(f))\triangleq \frac{1}{2^m}\max_{\biw\in(q/p)\Z_p^m}|F(\biw)|^2.
\eeq
If $p$ is omitted, we shall refer to the PAPR of $\Psi(f)$ under the classical ($2$-ary) WHT. It is apparent that
\beq
\PAPR_p(\Psi(f))\le\PAPR_q(\Psi(f)).
\eeq
We remark that this measure in fact arises from a more general definition of the PAPR under unitary transforms (cf. \cite{Parker2003}, \cite{Riera2006}). Also the PMEPR can be restated in this context \cite{Parker2003}. The PAPR under the $p$-ary WHT is an important measure in cryptographic applications \cite{Parker2003a}, and it is of interest in MC-CDMA communications systems \cite{Paterson2004}, particularly when $p=2$. 
\par
OFDM and MC-CDMA enjoy several similarities: in both cases codewords are used to modulate simultaneously a number of orthogonal signals, which are continuous in OFDM and discrete in MC-CDMA. In MC-CDMA systems the PAPR (under the classical WHT) turns out to be an analogous measure of the PMEPR in OFDM systems \cite{Paterson2002}. In \cite{Paterson2002} and \cite{Paterson2004} Paterson studied binary codes whose codewords have low PAPR, whereas constructing nonbinary codes is left as an open problem.
\par
Theorem \ref{thm:lower-bound} relates the PMEPR of cosets of $\RM_q(1,m)$ with the $q$-ary WHT of its coset representative. It is a consequence of Lemma \ref{lem:WHT-prop} that every word in a coset of $\RM_q(1,m)$ has the same PAPR under the $q$-ary WHT (but generally not under the $p$-ary WHT). Using this fact, Theorem \ref{thm:lower-bound} states that cosets with low PMEPR must have also low PAPR under the $q$-ary WHT, and therefore, also under the $p$-ary WHT. Thus we have established the following corollary.
\begin{corollary}
\label{cor:PAPR-WHT}
Let $f:\Z_2^m\rightarrow\Z_q$ be a generalized Boolean function, and suppose that $p$ divides $q$. Then the PAPR under the $p$-ary WHT of the polyphase codewords in the coset $\psi(f)+\RM_q(1,m)$ is at most the PMEPR of the polyphase codewords in the coset.
\end{corollary}
\par
The above corollary shows that all codes with low PMEPR that are unions of cosets of $\RM_q(1,m)$ (e.g., codes that arise from the results in the present paper and the codes constructed in \cite{Davis1999} and \cite{Paterson2000a}) also enjoy low PAPR. Hence with Corollary \ref{cor:PAPR-WHT} a number of (generally nonbinary) coding options for MC-CDMA with low PAPR can be derived. However, as opposed to many of the codes in \cite{Paterson2002} and \cite{Paterson2004} having PAPR equal to $1$, Corollary \ref{cor:PAPR-WHT} in combination with known results (e.g., Corollary \ref{cor:PMEPR_cosets}) yields an upper bound on the PAPR that is at least $2$. 
\par
We refer to the final section of \cite{Paterson2004} for further discussions on the connection between codes for OFDM and MC-CDMA.


\section{Construction Examples}
\label{sec:examples}

In the following we will apply Corollary \ref{cor:PMEPR_cosets} in order to find families of cosets of $\RM_q(1,m)$ with low PMEPR. By Corollary \ref{cor:PAPR-WHT} such cosets also have low PAPR under the $p$-ary WHT, where $p$ is a divisor of $q$. Trivially we may choose our kernel functions $a$ and $b$ from the complete set of generalized Boolean functions in $k$ variables. In such a way, we can construct large sequence families with their PMEPR bounded by $2^{k+1}$, since the trivial bound states \beq
\frac{\Phi(a)\star\Phi(b)}{2^k}\le2^{k+1}.
\eeq
However it can be shown that these sequence sets are just subsets of a larger family of complementary sets of size $2^{k+1}$ as constructed in \cite{Schmidt2006b} and \cite{Paterson2000a}. Hence this approach does not yield any new sequence sets.

\subsection{Sequence Sets with PMEPR at most 2}

An immediate consequence of Corollary \ref{cor:PMEPR_cosets} (and in this case also of Theorem \ref{thm:PMEPR_path}) is the well-known construction of Golay sequences over $\Z_q$, which has been stated by Davis and Jedwab in \cite{Davis1999} for $q$ being a power of 2. 
\begin{corollary}
\label{cor:DavisJedwab}
Each of the $m!/2$ cosets of $\RM_q(1,m)$ with coset representatives corresponding to
\beqn
\label{eqn:coset-2}
\frac{q}{2}\sum_{i=0}^{m-2}x_{\pi(i)}x_{\pi(i+1)},
\eeqn
where $\pi$ is a permutation of $\{0,1,\cdots,m-1\}$, is entirely comprised of Golay sequences. In particular the maximum PMEPR of the polyphase words in these cosets is 
(i) exactly $2$ if $m$ is odd and $q$ arbitrary (even) and if $m$ is even and $q\equiv 0\pmod 4$, 
(ii) at least $1+\cos^2\frac{\pi}{q}$ and at most 2 if $m$ is even and $q\equiv 2\pmod 4$.
\end{corollary}
\begin{proof}
The corollary follows by setting $k=0$, $a=0$, and $b=0$ in Corollary \ref{cor:PMEPR_cosets} and Theorem \ref{thm:LB-constr}. Since $\Phi(a)\star\Phi(b)=2$, the upper bound on the PMEPR is an immediate consequence of Corollary \ref{cor:PMEPR_cosets}. Note that $a$ and $b$ have $q$-ary WHT equal to $1$. For odd $m$ the lower bound on the PMEPR follows then immediately from (\ref{eqn:lower_bound_m_odd}). For even $m$ and $q\equiv 0\pmod 4$ put 
\beq
\biw=\left(\frac{q}{4},\,\frac{q}{4}\right)
\eeq
in the maximum calculation in (\ref{eqn:lower_bound_m_even}), which leads to a lower bound of $2$. For even $m$ and $q\equiv 2\pmod 4$ it is not hard to see that the maximum in (\ref{eqn:lower_bound_m_even}) is attained for
\beq
\biw=\left(\frac{q-2}{4},\,\frac{q+2}{4}\right).
\eeq
Hence, with $j=\sqrt{-1}$, (\ref{eqn:lower_bound_m_even}) becomes
\begin{align*}
\frac{1}{4}\max_{\biw\in\Z_q^2}\big|(1+\xi^{w_1})+\xi^{w_0}(1-\xi^{w_1})\big|^2
&=\frac{1}{4}\left|(1+je^{j\pi/q})+je^{-j\pi/q}(1-je^{j\pi/q})\right|^2\\
&=\frac{1}{4}\left|2+2j\Re\left\{e^{j\pi/q}\right\}\right|^2\\
&=1+\cos^2\frac{\pi}{q},
\end{align*}
which completes the proof.
\end{proof}
\par
We remark that the upper bound is well known: it was first proved by Davis and Jedwab for $q$ being a power of 2 \cite[Corollary 6, Corollary 9]{Davis1999} and generalized to arbitrary $q$ by Paterson \cite[Corollary 11]{Paterson2000a}. The lower bound in Corollary \ref{cor:DavisJedwab} (i) has been first proved Cammarano and Walker \cite{Cammarano1999}, and the case when $m$ is odd also by Paterson \cite[Theorem 21]{Paterson2000a}. Except for the trivial bound $1$ when $q=2$, the lower bound in Corollary \ref{cor:DavisJedwab} (ii) is new. Note that these upper and lower bounds on the PMEPR now arise in a uniform way from a general framework. From the discussion following Theorem \ref{thm:lower-bound} we also conclude that the considered cosets are optimal at least for $q=2$ and odd $m\le 7$ in the sense that for these parameters there cannot exist cosets with PMEPR lower than $2$.
\par
Li and Chu \cite{Li2005} reported 1024 additional quaternary (with elements in $\{1,-1,\sqrt{-1},-\sqrt{-1}\}$) Golay sequences of length 16. While Fiedler and Jedwab \cite{Fiedler2006} recently provided an explanation for these sequences, it has been observed earlier by Holzmann and Kharaghani \cite{Holzmann1994} that there exists essentially one quaternary kernel of length 8 
\beq
((+ + + - + + - +)\;,\; (+ \;j\;j - + \;i\;i -)),\qquad j=\sqrt{-1},i=-j.
\eeq
These two sequences are associated with generalized Boolean functions $\Z_2^3\rightarrow\Z_4$ corresponding to the forms
\beq
2x_0x_1+2x_1x_2\qquad\mbox{and}\qquad 2x_0x_2+2x_1x_2+x_0+x_1,
\eeq
respectively. Following \cite{Holzmann1994} from this kernel 512 equivalent kernels can be generated by applying several operations that preserve the complementary property. These kernels can be used in conjunction with Theorem \ref{thm:PMEPR_path} to obtain an explicit construction for Golay sequences of length $2^m$, where $m>3$. Notice that all of them correspond to cubic forms. Indeed, if $m=4$, we obtain exactly the 1024 Golay sequences of length 16 reported by Li and Chu \cite{Li2005}. Unfortunately, in this way, we cannot construct complete cosets of $\RM_q(1,m)$ whose PMEPR is bounded by 2. Instead Corollary \ref{cor:PMEPR_cosets} reveals that the PMEPR is at most 5 if the 512 kernels are used in conjunction with Corollary \ref{cor:PMEPR_cosets} to build cosets of $\RM_q(1,m)$.

\subsection{Sequence Sets with Maximum PMEPR between 2 and 4}
\label{sec:PMEPR24}
Consider the functions $a,b:\Z_2^2\rightarrow\Z_q$ with
\begin{align*}
a=a(x_0,x_1)&=\frac{q}{2} x_0x_1\\
b=b(x_0,x_1)&=\frac{q}{2} x_0x_1+\left(\alpha +\frac{q}{2}\right)x_0+\beta x_1,
\end{align*}
where $\alpha,\beta\in(q/p)\Z_p$ and $p$ divides $q$. We have
\begin{align*}
\Phi(a)&=(1\;\;1\;\;0\;\;0\;\;1\;\;-1)\\
\Phi(b)&=(1\;\;-\xi^{\alpha}\;\;0\;\;0\;\;\xi^\beta\;\;\xi^{\alpha+\beta})
\end{align*}
and
\begin{align*}
(A(\Phi(a))(\ell)\,|\,0\le\ell<6)&=(4\;\;0\;\;0\;\;1\;\;0\;\;-1)\\
(A(\Phi(b))(\ell)\,|\,0\le\ell<6)&=(4\;\;0\;\;0\;\;-\xi^{\beta-\alpha}\;\;0\;\;\xi^{\beta+\alpha}).
\end{align*}
Hence
\beq
\Phi(a)\star\Phi(b)=2(4+|1-\xi^{\beta-\alpha}|+|1-\xi^{\beta+\alpha}|)\le 16.
\eeq
So, by Corollary \ref{cor:PMEPR_cosets}, the PMEPR of the cosets of $\RM_q(1,m)$ ($m>2$) with coset representatives corresponding to
\beqn
\label{eqn:coset-alpha-beta}
\frac{q}{2}\sum_{i=0}^{m-2}x_{\pi(i)}x_{\pi(i+1)}+\alpha x_{\pi(0)}x_{\pi(2)}+\beta x_{\pi(1)}x_{\pi(2)},
\eeqn
is at most $4$. These cosets lie inside $\RM_q(2,m)$ and in particular inside $\ZRM_q(2,m)$ if $q>p$.
\par
There are $p^2$ choices for the pair $\alpha,\beta$ and for each pair we can construct $m!/2$ distinct cosets by applying to the variable indices $m!/2$ permutations that are invariant under reversal. However setting $\alpha=\beta=0$ or $\alpha=\beta=q/2$ in (\ref{eqn:coset-alpha-beta}) will generate the same set of coset representatives. Hence the total number of distinct coset representatives amounts to $(p^2-1)(m!/2)$. 
\par
For particular values for $\alpha$ and $\beta$ the upper bound on the PMEPR of the cosets can be tightened. If $\alpha=\beta=0$ or $\alpha=\beta=q/2$, we obtain sequences with PMEPR at most 2. However in this case the coset representatives are of the form given in (\ref{eqn:coset-2}). More generally, the $(p^2-1)(m!/2)$ cosets of $\RM_q(1,m)$ organize in classes with the same PMEPR upper bound. For $p=2$ we obtain just two classes: $m!/2$ cosets with maximum PMEPR 2 and $m!$ cosets with PMEPR at most 4. Using Theorem \ref{thm:LB-constr} it can be shown that the latter bound is tight if $q\equiv0\pmod 4$. 
\par
For $q\ge4$ and $p=4$ we have $\alpha,\beta\in\left\{0,\frac{q}{4},\frac{q}{2},\frac{3q}{4}\right\}$ and the classes are:
\begin{center}
\begin{tabular}{c|c|c|c}
$\beta+\alpha$ & $\beta-\alpha$ & max. PMEPR & \# cosets\\ \hline
0   & 0     & 2  & $m!/2$\\
arbitrary (0)  & 0 (arbitrary)  & 3  & $2m!$\\
$\frac{q}{4}\pmod {\frac{q}{2}}$ & $\frac{q}{4}\pmod {\frac{q}{2}}$ & $2+\sqrt{2}$  & $4m!$\\
arbitrary   & arbitrary     & 4 & $m!$\\
\end{tabular}
\end{center}
Notice that the number of cosets are counted by considering only those cosets that are not contained in a class with a lower PMEPR bound. For example the total number of cosets with PMEPR at most 3 is equal to $5m!/2$, while there are $13m!/2$ cosets with PMEPR at most $(2+\sqrt{2})$. Employing Theorem \ref{thm:LB-constr} it can be shown that the PMEPR bounds given above are tight if $q\equiv0\pmod 8$. If $q=4$, Theorem \ref{thm:LB-constr} yields a lower bound of $2$ for the first class, $2.5$ for the second and third class, and $4$ for the fourth class.
\par
We remark that the second class includes those $48$ cosets of $\RM_{8}(1,4)$ inside $\ZRM_{8}(2,4)$ for which Davis and Jedwab observed that their PMEPR is exactly 3 \cite[Tables V and VI]{Davis1999}. While Nieswand and Wagner \cite{Nieswand1998} could explain this behaviour partially, a complete proof arises as a special case of Corollary \ref{cor:PMEPR_cosets} and Theorem \ref{thm:LB-constr}\footnote{The upper bound for this particular case has been proved alternatively by the author and Finger in \cite{Schmidt2006b}.}. Moreover we identified further cosets of $\RM_q(1,m)$ with PMEPR at most 3. 
\par
For $q\ge8$ and $p=8$ there are already 9 classes with different PMEPR upper bounds. These are:
\begin{center}
\begin{tabular}{rl|c}
\multicolumn{2}{c|}{max. PMEPR} & \# cosets\\ \hline
&2 & $m!/2$\\
$2+1/\sqrt{2}\;\;\;\approx$&2.707 & $4m!$\\
$2+\sqrt{2-\sqrt{2}}\;\;\;\approx$&2.765 & $4m!$\\
&3 & $2m!$\\
$2+\sqrt{1+1/\sqrt{2}}\;\;\;\approx$ & 3.307 & $8m!$\\
$2+\sqrt{2}\;\;\;\approx$ & 3.414 & $4m!$\\
$3+1/\sqrt{2}\;\;\;\approx$ & 3.707 & $4m!$\\
$2+\sqrt{2+\sqrt{2}}\;\;\;\approx$ & 3.848 & $4m!$\\
& 4 & $m!$\\
\end{tabular}
\end{center}
Applying Theorem \ref{thm:LB-constr} we are able to show that the PMEPR bounds are tight if $q\equiv 0\pmod {16}$.
\par
The number of suitable kernels of length 4 producing cosets of $\RM_q(1,m)$ with maximum PMEPR between $2$ and $4$ can be increased further. To this end, consider the pairs $(a,b)$, where $a,b:\Z_2^2\rightarrow\Z_q$ are given by
\begin{align*}
a=a(x_0,x_1)&=\gamma x_0x_1\\
b=b(x_0,x_1)&=\delta x_0x_1+\left(\alpha +\frac{q}{2}\right)x_0+\beta x_1,
\end{align*}
$\alpha,\beta,\gamma,\delta\in\Z_q$, and $\gamma$ and $\delta$ are selected such that $\Phi(a)\star\Phi(b)\le 16$. However notice that the coset representatives of the constructed cosets of $\RM_q(1,m)$ generally correspond to cubic forms.
\par
It is noteworthy that a computer-based search failed to find kernels that generate more cosets of $\RM_q(1,m)$ with $\PMEPR$ less than 4. Our search included all kernel functions in $k$ variables for $k=4$ and $q=2$, $k=3$ and $q=4$, and quadratic functions with $k=5$ and $q=2$. We leave the identification of more cosets of $\RM_q(1,m)$ with low PMEPR (preferably close to 2) to further work.

\section{Concluding Remarks}
\label{sec:conclusion}

In this paper we have shown how cosets of $\RM_q(1,m)$ with low PMEPR can be obtained in a systematic way once suitable kernels are known. The PMEPR of these cosets is low because such cosets are comprised of sequences lying in so-called near-complementary pairs. It was demonstrated that those cosets of $\RM_q(1,m)$ comprised entirely of Golay sequences simply arise as a special case in a more general theory. We presented some suitable kernels and, in this way, several previously unexplained phenomena arising in earlier works by Davis and Jedwab \cite{Davis1999}, by Paterson \cite{Paterson2000a}, and by Parker and Tellambura \cite{Parker2001b}, \cite{Parker2001} can now be understood in a general framework. Moreover we provided at least a partial answer to the question stated in \cite{Paterson2000a}: {\it What other regularities appear in the PMEPRs of cosets as we move to higher alphabets, and how can they be explained in general?} We have also established a connection between the OFDM and MC-CDMA coding problems, and suggested (generally nonbinary) coding solutions for MC-CDMA with low PAPR.
\par
We wish to point out some relations to the work by Parker and Tellambura \cite{Parker2001b}, \cite{Parker2001}. Indeed
Theorem \ref{thm:PMEPR_path} in the present paper is similar to \cite[Theorem 5]{Parker2001} and \cite[Theorem 6]{Parker2001b}, which also exploit the Rudin--Shapiro construction. However these references do not identify the crucial connection between the PMEPR and the aperiodic auto-correlation, which allows us to extend the PMEPR bound to a slightly weaker bound in Theorem \ref{thm:PMEPR_coset} and Corollary \ref{cor:PMEPR_cosets} that holds for complete cosets of $\RM_q(1,m)$. In contrast, in order to obtain cosets of $\RM_2(1,m)$ with low PMEPR, the approach in \cite{Parker2001b} and \cite{Parker2001} involves a computational search for the maximum PMEPR over a number of kernels. Indeed in \cite[Table 6]{Parker2001} it is implicitly shown that the cosets with coset representatives given in (\ref{eqn:coset-alpha-beta}) have PMEPR at most 4 when $q=2$. Notice that their semi-computationally obtained upper bound is exactly as predicted by Corollary \ref{cor:PMEPR_cosets}.
\par
We presented some kernels of short length and we also sought good kernels of medium length. From the complexity point of view it is also realistic to perform an exhaustive search for good kernels of larger lengths, say 32 or 64, and defined over larger alphabets. We have not attempted such a search and leave it to further work. It would be desirable to find kernels producing cosets of $\RM_q(1,m)$ with PMEPR close to 2, or even with PMEPR of exactly 2 (which implies the discovery of new complementary pairs). Moreover our theory would benefit from having an efficient way to construct good kernels.


\end{document}